\documentclass[fleqn,twoside]{article}
\usepackage{espcrc2}

\usepackage{graphicx}
\usepackage{epsfig}

\newcommand{\AmS}{{\protect\the\textfont2
  A\kern-.1667em\lower.5ex\hbox{M}\kern-.125emS}}

\newcommand{\eq}{\begin{equation}}
\newcommand{\en}{\end{equation}}
\newcommand{\eqa}{\begin{eqnarray}}
\newcommand{\ena}{\end{eqnarray}}
\def\Journal#1#2#3#4{{#1}{\bf #2} (#4) #3}

\def\PLB{Phys. Lett.~{\bf B}}

\def\PRD{Phys. Rev. ~{\bf D}}

\begin{document} 
\title{
\vspace{-35mm}
\rightline{\small RNCP-Th01026~~~~~~}
\rightline{\small HU-EP-01/44~~~~~~~~}
\vspace{10mm}
Classical solutions with nontrivial holonomy 
in $SU(2)$ LGT at $T \ne 0$%
\thanks{Poster presented by A.I. Veselov at LATTICE 2001, Berlin; 
extended hep-lat version of the proceedings contribution.}}
\author{E.-M. Ilgenfritz%
\address{Research Center for Nuclear Physics, Osaka University,
Osaka 567-0047, Japan}
\thanks{E.-M.I. gratefully appreciates the support by the Ministry
of Education, Culture and Science of Japan (Monbu-Kagaku-sho) and
thanks for a CERN visitorship.},
B. V. Martemyanov\address[ITEP]{Institute for Theoretical and Experimental 
Physics, Moscow 117259, Russia}
\thanks{This work was partly supported by RFBR grants 99-01-01230 
and 01-02-17456, and INTAS 00-00111.},
M. M\"uller-Preussker\address{Humboldt-Universit\"at zu Berlin, 
Institut f\"ur Physik, 10115 Berlin, Germany} and
A. I. Veselov\addressmark[ITEP]
}

\begin{abstract}
  We generate $SU(2)$ lattice gauge fields at finite temperature 
  and cool them in order to characterize the two phases by the 
  occurrence of specific classical solutions.   
  We apply two kinds of spatial boundary conditions: fixed holonomy  
  and standard periodic b.c. 
  For $T < T_c$ our findings concerning classical configurations 
  semi-quantitatively agree for both types of boundary conditions.
  We find in the confinement phase a mixture of undissociated
  calorons with lumps of positive or negative half-integer topological 
  charges.   
\end{abstract}

\maketitle

\setcounter{footnote}{0}

\section{INTRODUCTION}
There is one scenario of confinement which envisions the ground
state of QCD as a dual superconductor formed by the condensation of 
abelian magnetic monopoles. Nowadays links are established to a
vortex condensation picture of confinement.
Still, one might be puzzled by the question which role the carrier 
of topological charge (easily associated with chiral symmetry breaking) 
could play for confinement. Dealing with dilute, uncorrelated gases of 
instantons \cite{BPST} 
or - at finite temperature - of calorons \cite{HS} 
cannot explain confinement. Abandoning the tacit assumption of trivial 
holonomy at spatial infinity has opened the door to a larger class of 
selfdual solutions at finite temperature \cite{VB}. The new calorons 
with nontrivial holonomy  boundary conditions are interesting in the 
light of the riddle because, generically, they are composed of magnetic 
monopoles (dyons). Configurations of this type have been studied both 
analytically (in continuum) and numerically (in $SU(2)$ lattice gauge 
theory). The fixed holonomy conditions \cite{PVB,IMV} have been mimicked 
on the lattice by fixing the temporal links everywhere on the {\it spatial} 
surface ($i_x=L_x$ etc.) to a constant $U_4$ such that the Polyakov line 
$L({\vec x})$, the trace of the holonomy ${\cal P}({\vec x})= 
\prod_{t=1}^{L_4}~U_{{\vec x},t,4}$,
is put equal to the statistical average $\langle L \rangle$ at the 
given temperature (see \cite{IMMV}). For the $SU(2)$ case a clear 
correspondence between analytical solutions and lattice configurations 
has been established \cite{IMMV}. These lattice studies have copiously 
produced also (approximate) solutions, composed of {\it opposite sign} 
approximately half-integer topological charges ($D\bar D$ pairs) \cite{IMMV}.
They do not correspond to any known analytical solution. In this paper we
report on results showing that approximately $O(4)$ symmetric calorons 
(denoted $CAL$) as well as static $DD$ and $D\bar D$ 
pairs are statistically significant in the confinement phase ($0 < T \le T_c$).
In this case, fixed holononomy would require to put $L=0$ on the boundary.
However, our observations show no qualitative difference to periodic boundary 
conditions {\it without} this constraint.
\vspace{-3mm} 

\section{CLASSICAL CONFIGURATIONS}

We consider again pure $SU(2)$ gluodynamics with Wilson action
at finite temperature, but this time comparing two types of spatial 
boundary conditions, ({\it i}) periodic b.c. with time-like link 
variables being fixed on the spatial surface and ({\it ii}) standard 
periodic b.c. without constraints. 
We mostly used a lattice of size $16^3 \times 4$, in a few cases also 
$32^3 \times 4$. We took coupling values $\beta$ below and above 
the deconfinement transition ($\beta_c \simeq 2.29$) to create equilibrium 
configurations to start with. The same b.c. are applied for the Monte Carlo 
process and for cooling. 

In a first stage we have searched for topologically non-trivial objects with 
lowest possible action late in the cooling history in order to find systematic 
dependences on the phase and the boundary conditions. Cooling was stopped 
at the $n$-th cooling step when the following criteria were fulfilled: 
action  $S_n < 2~S_{\mathrm{inst}}$, change of action $|S_n - S_{n-1}| < 0$
and concavity $S_n-2~S_{n-1}+S_{n-2} < 0$, {\it i.e.} cooling just passed 
a point of inflection.\footnote{$S_{\mathrm{inst}}=2 \pi^2 \beta~$ 
is the action  of a single instanton.} 
For each $\beta$ we have scanned the topological 
content of $O(200)$ configurations. In this late stage we find approximate 
classical solutions, more or less static. We define the ''non-staticity'' 
in (Euclidean) time by 
$$T_s= \sum_i |S_i - S_{i-1}| / \sum_i S_i \, , $$
where $S_i$ denotes the action in the $i$-th timeslice.
According to Refs. \cite{IMV,IMMV}, the almost-classical configurations
can be classified as $DD$, $CAL$, $D\bar D$, and purely magnetic (Dirac) 
single and double sheets - $M$ and $2M$.
Some configurations undergoing cooling do not match the stopping criteria. 
They turn into trivial vacuum states.

Calorons ($CAL$) are a limiting case of $DD$ configurations \cite{PVB,IMMV}. 
The latter have opposite sign peaks of $L({\vec x})$ near the centers of the 
lumps of action and topological charge. If the peaks of topological charge 
are too close to be separated in $3D$ space, these objects happen to be 
non-static as well. For the confinement phase we have monitored how 
frequently objects with given $T_s$ are found. The histograms are rather 
similar for both types of b.c. They have a peak at $T_s= 0.02-0.04$ and 
have a long tail of widely varying $T_s$. To enable an easy distinction between
$DD$ and (non-ideal) $CAL$ events we have searched and verified a cut in 
$T_s$. For $T_s < 0.17$ we may classify the objects as $DD$ (static with two 
well-separated maxima of the densities of topological charge $q({\vec x})$ 
and action $s({\vec x})$). For $T_s > 0.17$ the objects can be classified as 
$CAL$ (non-static, approximately $O(4)$ symmetric solutions, with a single 
maximum of the $3D$ projected $q({\vec x})$ and $s({\vec x})$).
$D\bar D$~ have been found always to consist of two well-separated static 
objects with nearly vanishing $T_s = 0.004 \pm 0.002$, much smaller than 
for $DD$'s ! Purely magnetic sheets - $M$ have $S_{\mathrm{magnetic}} >> 
S_{\mathrm{electric}}$, with an action quantized in units of 
$S_{\mathrm{inst}}/4$. They are also perfectly static with 
$T_s = 0.003 \pm 0.002$.

As can be seen from Table 1, the relative frequency to obtain different types
of nearly classical configurations ($DD$, $CAL$, $D\bar D$, $M$ and $2M$) 
is quite different, depending on whether cooling starts from Monte Carlo 
configurations in the confinement phase or the deconfinement phase. 

\vspace{-6mm}
\begin{table}[h]
\caption{Relative frequencies of 
cooled quasi-stable configurations 
for $\beta=2.2, 2.25$ (confinement phase) and 2.35 (deconfinement phase. 
First and second rows
refer to fixed holonomy and standard periodic b.c.,
respectively. The lattice size is $16^3 \times 4$.}
\vspace{3mm}
\begin{tabular}{lccc} 
\hline 
Type& $\beta=2.20$&$\beta=2.25$&$\beta=2.35$ \\
\hline
$DD$ 
     & $.46 \pm .05$&$.52 \pm .05$&$.20 \pm .03$ \\
     & $.43 \pm .05$&$.44 \pm .05$&$.01 \pm .01$  \\ 
$CAL$
     & $.19 \pm .03$&$.17 \pm .03$&$.04 \pm .01$ \\
     & $.24 \pm .03$&$.26 \pm .03$&$.06 \pm .02$  \\ 
$D\overline{D}$ 
     & $.28 \pm .04$&$.26 \pm .04$&$.58 \pm .05$  \\
     & $.18 \pm .03$&$.16 \pm .03$&$0.0$          \\ 
$M$, $2M$   
     & $.01 \pm .01$&$.01 \pm .01$&$.10 \pm .02$  \\
     & $.04 \pm .02$&$.03 \pm .01$&$.22 \pm .03$  \\ 
trivial vac.
     & $.06 \pm .02$&$.04 \pm .02$&$.08 \pm .02$  \\
     & $.11 \pm .02$&$.11 \pm .02$&$.71 \pm .06$  \\ 
\hline
\end{tabular}
\end{table}
\vspace{-6mm}

For the confinement phase we find that the relative probabilities for all 
objects are approximately independent of the type of b.c.
For the deconfinement phase we see that the strong enhancement of $D\bar D$ 
configurations earlier found for fixed holonomy b.c. \cite{IMMV} (which 
would be compatible with the suppression of the topological susceptibility) 
is {\it not} reproduced for standard periodic b.c. In the standard case, the 
probability to obtain any topologically non-trivial object drops sharply 
with $\beta$.
As for the deconfinement phase as such, the latter observation must be 
considered with caution since the physical 3-volume is very small. 
The independence of the boundary conditions, however, in the confinement
phase must be taken seriously: the enforcement of $L=0$ boundary
conditions seems to be equivalent with the conditions under normal 
(thermal) boundary conditions.

\section{DILUTE GASES AT HIGHER ACTION}

Applying the $L=0$ fixed holonomy b.c. we have studied in more detail  
the configurations closer to equilibrium in the confinement phase
which represent snapshots of the early cooling history.
The stopping criterion above (however, without limitation of action)
has been applied and the configurations have been automatically stored.
This gives a series of subsequent ''plateaux'' of the action, while
the cooling history in between is always reproducible.
In terms of the objects classified above, we have scanned the 
topological content of subsequent plateau confi\-gurations.
In the first plateau we find an uncorrelated gas of dyons $D$ and
antidyons $\bar D$ carrying roughly half-integer topological charges. 
We identified the dyons simultaneously by various measurable quantities: 
extrema of the Polyakov line $L({\vec x})$, of the topological 
charge and action density and of the non-Abelianicity (after the maximal 
Abelian gauge
has been fixed) and finally by the static Abelian monopoles (defined in Abelian 
projection).
Their world lines coincide with the world lines of dyons. If the signs of
the monopole charge of an Abelian monopole and of $L({\vec x})$ (measured 
in the dyon center ${\vec x}$) are the same (opposite), the topological 
charge of the dyon is approximately equal to $+\frac{1}{2}$ (or $-\frac{1}{2}$).

The cooling trajectory between this stage and the final stage discussed 
in the previous section is a series of annihilations. Between subsequent
plateaux only such annihilation events of topological lumps have been 
observed which are monopole-antimonopole annihilations (conservation of
magnetic charge). Depending on the re\-lative sign of $L({\vec x})$ in the 
center, these are $DD$ or $D\bar D$ annihilations. 
The two cases are shown in the Figure.
Neither electric nor topological charge is conserved under (Wilson) cooling. 
In $D\bar D$ annihilations a sharp drop of the topological charge density 
can be seen while the action density and Polyakov line slowly relax to zero. 
$DD$ annihilations resemble how $T=0$ lattice instantons disappear
under cooling with Wilson action: topological charge density and action 
density increase, turning everything into a singular ''dislocation'' 
which finally collapses. At the end of the cooling process, 
when $S = O(1)~S_{\mathrm{inst}}$, we find either $DD$ (incl. $CAL$) 
or $D\bar D$ configurations.
The first ones (with $Q=\pm1$) can decay as described, while $D\bar D$
annihilate into $M$, $2M$ configurations or to the trivial vacuum. 

\begin{figure}[!htb]
\begin{center}
(a)\includegraphics[width=0.5\textwidth]{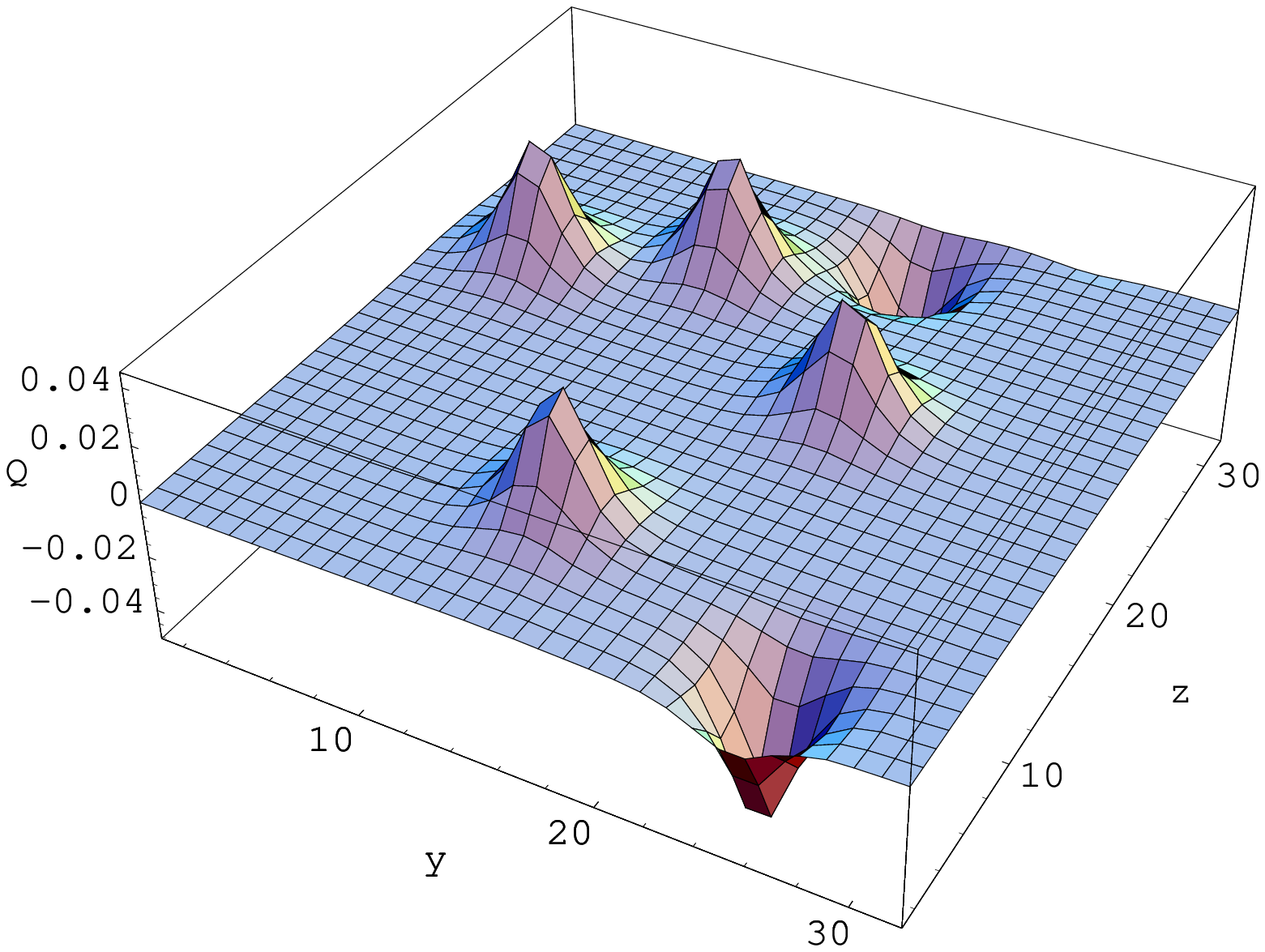}%
(a')\includegraphics[width=0.5\textwidth]{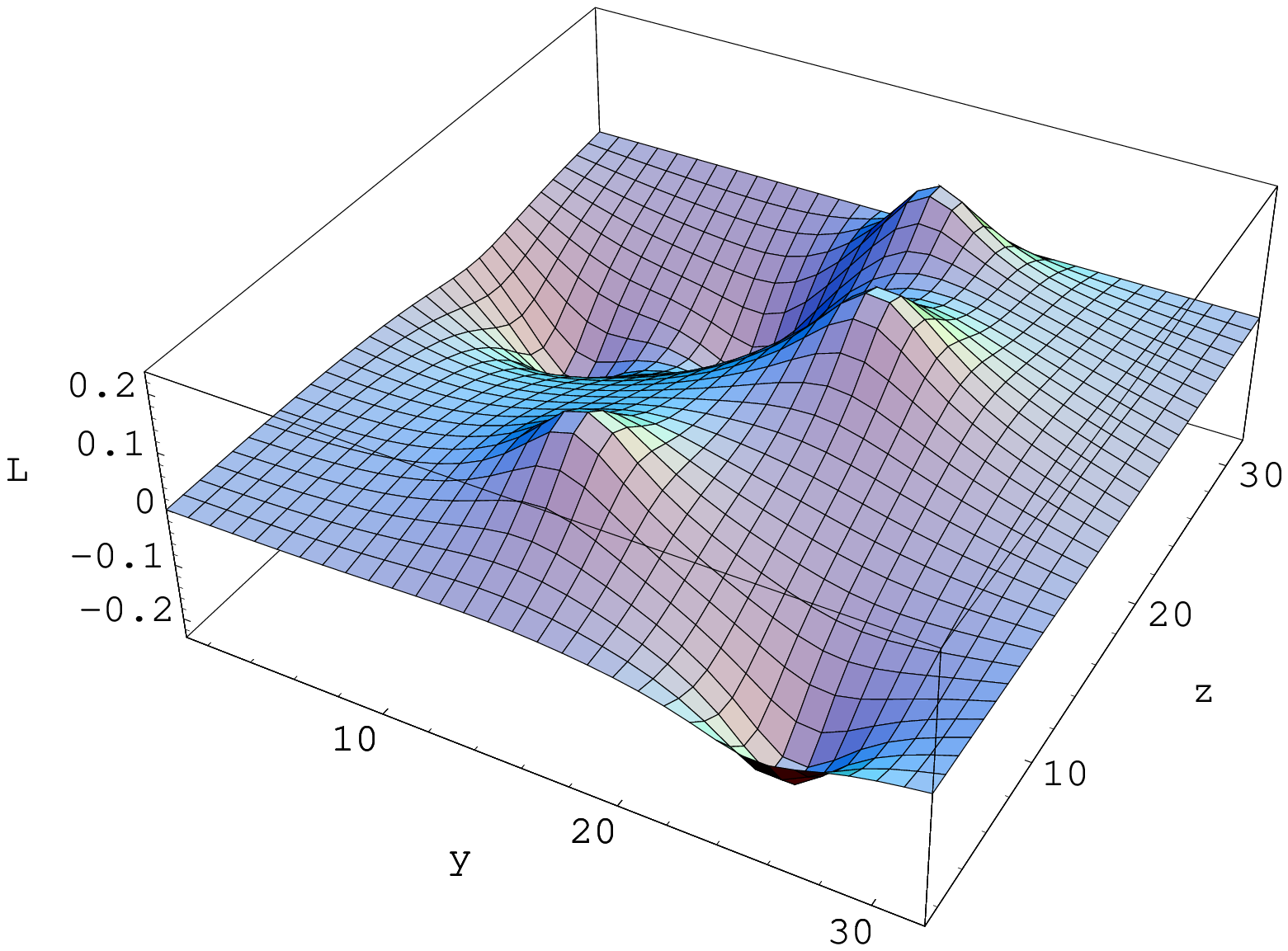}

(b)\includegraphics[width=0.5\textwidth]{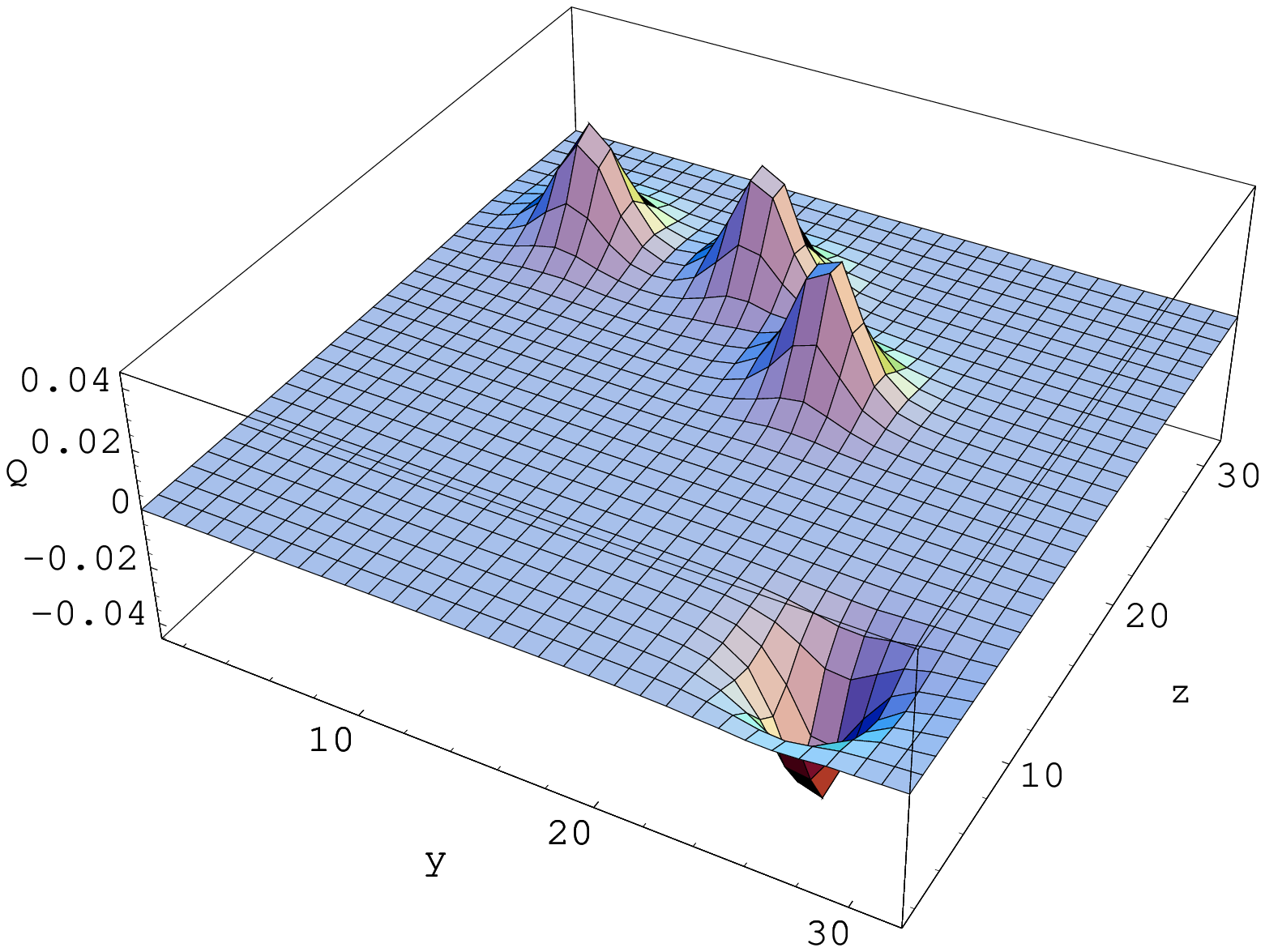}%
(b')\includegraphics[width=0.5\textwidth]{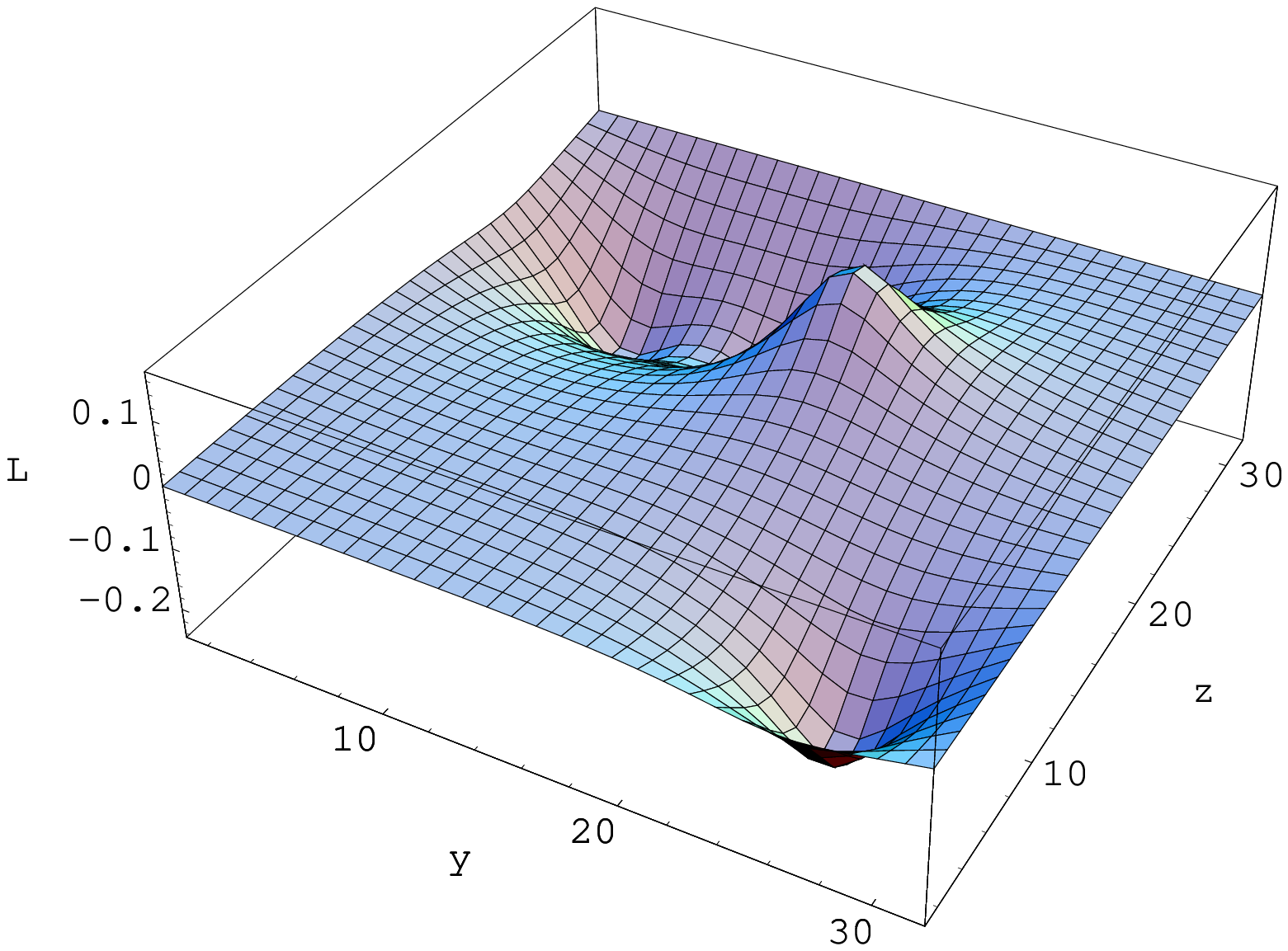}
(c)\includegraphics[width=0.5\textwidth]{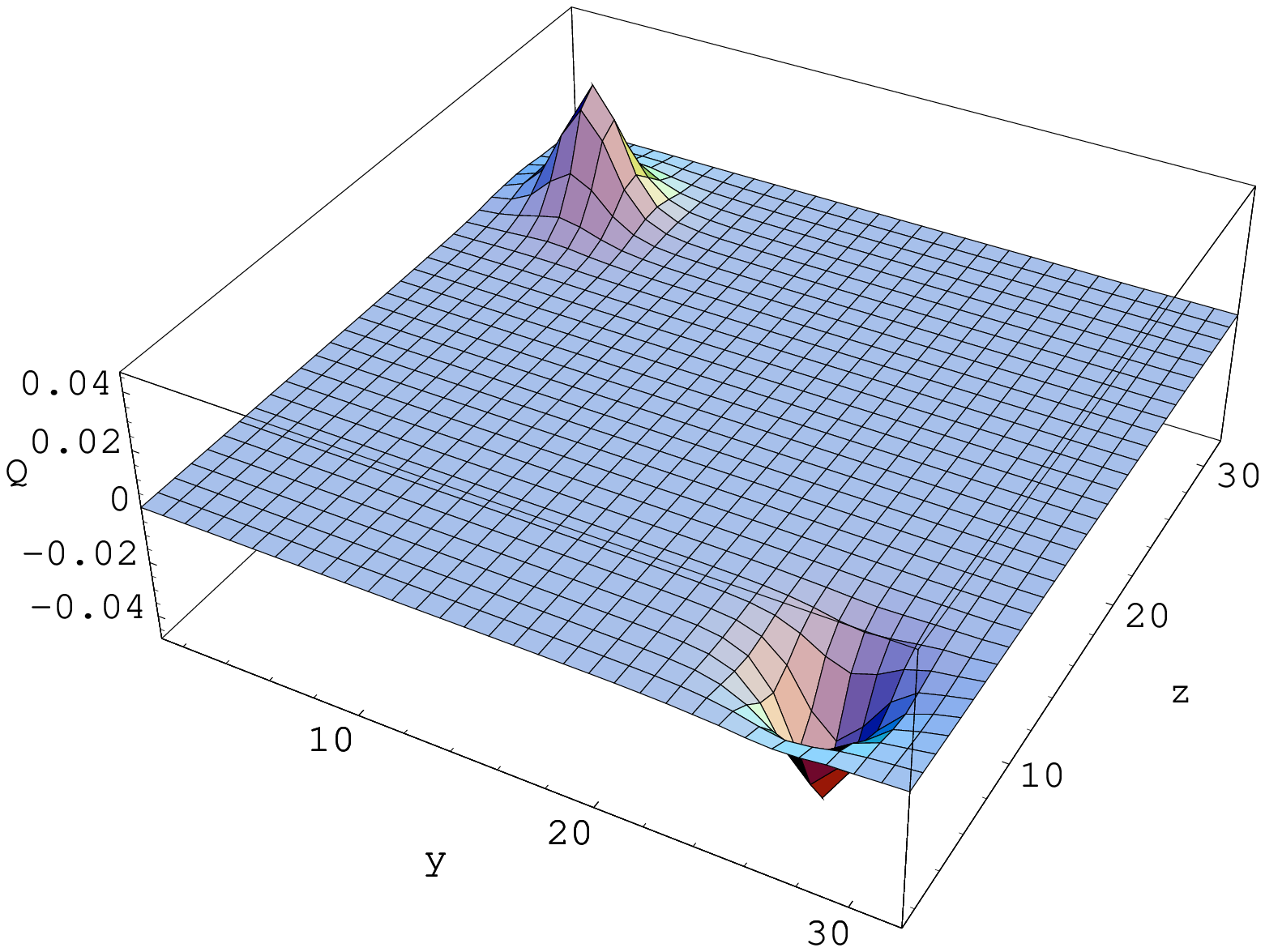}%
(c')\includegraphics[width=0.5\textwidth]{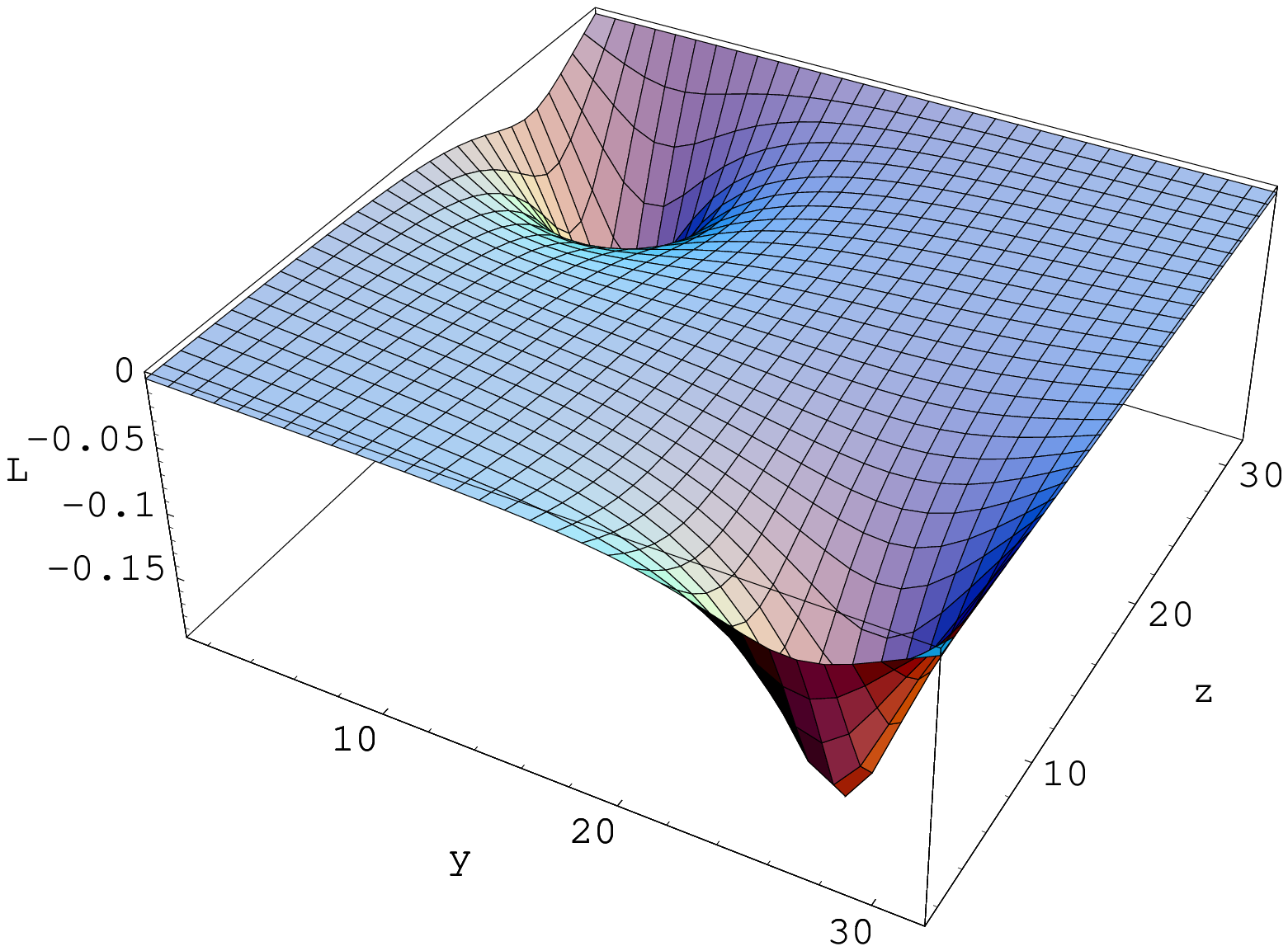}
\label{an}
\caption{
Topological charge density (a, b, c) and corresponding spatial 
Polyakov line distribution (a', b', c') at different cooling 
stages for a typical gauge field configuration. The transition
(a, a') $\rightarrow$ (b, b') shows the annihilation of a 
$D\bar D$ pair and (b, b') $\rightarrow$ (c, c') 
the annihilation of a $DD$ pair, respectively.
}
\end{center}
\end{figure}

\section{SUMMARY}
Calorons with non-trivial holonomy have motivated us to reconsider the 
topological content of the vacuum at finite temperature by careful cooling.
We cannot claim, that finite temperature gauge fields can be understood 
semiclassically with the new calorons as the only background.
First, with cooling also nearly classical configurations of dyon-antidyon 
type are obtained. Secondly, closer to the equilibrium, cooling with fixed 
holonomy boundary conditions produces a dilute gas of dyons and antidyons. 
For the confinement phase we have found that fixed holonomy and usual 
periodic b.c. have less influence on cooling and the detection of 
a semiclasical topological background.  

The authors are grateful to P. van Baal, S.V. Molodtsov, M.I. Polikarpov,
and Yu.A. Simonov for useful discussions.

\end{document}